\newcommand{\beq} {\begin{eqnarray}}
\newcommand{\eeq} {\end{eqnarray}}
\documentclass[12pt, preprint,preprintnumbers,amsfonts,aps,nofootinbib]{revtex4}

\begin{document}

\title{Boundary Entropy Can Increase Under Bulk RG Flow}

\author{Daniel Green}
\email{drgreen@stanford.edu}
\author{Michael Mulligan}
\email{mcmullig@stanford.edu}
\author{David Starr}
\email{dbstarr@stanford.edu}
\affiliation{SLAC and Department of Physics, Stanford University, Stanford, CA 94305-4060} 

\preprint{SU-ITP-07/19}
\preprint{SLAC-PUB-12907}

\newcommand{\bra}[1]{\left\langle \,#1\, \right|}
\newcommand{\ket}[1]{\left| \,#1\, \right\rangle}
\newcommand{\braket}[2]{\left\langle \,#1\, | \,#2\, \right\rangle}

\begin{abstract}
The boundary entropy $\log(g)$ of a critical one-dimensional quantum
system (or two-dimensional conformal field theory) is known to
decrease under renormalization group (RG) flow of the boundary theory.
We study instead the behavior of the boundary entropy as the
\emph{bulk} theory flows between two nearby critical points.  We use
conformal perturbation theory to calculate the change in $g$ due to a
slightly relevant bulk perturbation and find that it has no preferred
sign.  The boundary entropy $\log(g)$ can therefore increase during
appropriate bulk flows.  This is demonstrated explicitly in flows
between minimal models.  We discuss the applications of this result to
D-branes in string theory and to impurity problems in condensed
matter.
\end{abstract}
\maketitle


\section{Introduction}
Two-dimensional quantum field theories play an important role in many
different branches of physics, ranging from string theory to condensed
matter.  It is therefore important to understand the general features
and qualitative behavior of these theories.  Much is known when the
theory is conformally invariant (CFT), due to the large symmetry
group.  However, less is known about the general features of
non-conformal theories.  One means of gaining insight is to perturb a
given critical theory and follow the resulting trajectory under RG
flow.  In particular, it is interesting to learn what new critical
theory is the endpoint of the RG flow and to compare the properties of
the new and old systems.

A famous result in this direction is the Zamolodchikov $c$ theorem
\cite{Zamolodchikov:1986gt}, which states that the central charge of 
a CFT always decreases under such a flow.  There is an analogous
theorem which holds for two-dimensional CFTs with boundary (BCFTs).
For these systems Affleck and Ludwig
\cite{Affleck:1991tk} introduced a quantity $g$, known as the
generalized ``ground-state degeneracy'' or ``boundary entropy.''  The
$g$ theorem states that this quantity decreases under RG flow of the
boundary, so long as the bulk theory remains critical during the
boundary flow.  This theorem has been proven both perturbatively
\cite{PhysRevB.48.7297} and non-perturbatively \cite{Friedan:2003yc}.  
However, \textit{a priori} there is no reason for this qualitative
behavior of $g$ to persist when the bulk theory also undergoes RG
flow.

There are many situations where one is specifically interested in
quantities related to the boundary.  For example, in string theory the
boundary is associated with D-branes, and the brane tension is given
by $g$ \cite{Polchinski:1995mt,Harvey:1999gq}.  The Kondo effect can
be understood using a BCFT
\cite{PhysRevLett.68.1046,PhysRevB.48.7297,Affleck:1995ge} for which
$\log(g)$ is the impurity entropy.  These are physical quantities
which one might wish to track as the bulk theory moves between two
critical points.

In this note we consider the change in $g$ during RG flow between two
bulk critical points.  Rather than tackle the problem in its full
generality, we suppose the bulk flow is induced by a slightly relevant
bulk perturbation and the new fixed point is near the old one.  Thus
we may use conformal perturbation theory to calculate the change in
$g$.  It was a similar analysis which motivated the original $g$
theorem conjecture by Affleck and Ludwig
\cite{Affleck:1991tk}.  This method was also used in
perturbative proofs of the $c$ theorem \cite{Ludwig:1987gs} and the
$g$ theorem \cite{PhysRevB.48.7297}.

There are two good reasons to take this approach.  First, it avoids
the problem of finding an appropriate off-critical definition of $g$, because
one is always computing $g$ at a fixed point.  Some previous work
introduced a particular off-critical definition that is amenable to
calculation; however, this might not be the best criteria to
use \cite{Dorey:2004xk,Dorey:2005ak}.  Second, our result will depend
only on general features of the conformal field theory.  Previous work
on bulk flow in theories with boundaries has largely focused on
specific models
\cite{Ghoshal:1993tm,LeClair:1995uf,Lesage:1998qf,Ahn:1998xm,Dorey:1999cj,Dorey:2004xk,Dorey:2005ak,Green:2006ku,Gaberdiel:2007us}.
Because each model has very different properties, it is
difficult to identify the general features.  For example, the bulk
flow studied in \cite{Dorey:1999cj,Dorey:2004xk,Dorey:2005ak} focused
on the non-unitary Lee-Yang model.  Features of these flows could be
attributed to the lack of unitarity, rather than being generic to bulk
flows.  Nevertheless, in \cite{Dorey:2004xk,Dorey:2005ak}, for a given definition of $g$ away from the critical point, $g$ was seen to increase in some unitary models for parts of the flow.

We begin in section \ref{sec:g} with a brief overview of $g$.  Next,
in section \ref{sec:calc}, we calculate the change in $g$ under bulk
RG flow.  We find that the change in $g$ does not have a definite
sign, and so it can increase during certain bulk flows.  In section
\ref{sec:mm} we demonstrate this behavior explicitly for certain flows
between minimal models.  We then go on in section \ref{sec:interp} to
discuss various applications of this result.

\section{Overview of $\mathbf{g}$}
\label{sec:g}

Consider a one-dimensional quantum system of length $L$ at a
temperature $T=1/\beta$, with boundary conditions $A$ and $B$ at the
two ends.  Near criticality, the free energy $\log Z$
takes the form \cite{Bloete:1986qm, Affleck:1986bv}
\begin{equation}\label{eq:defineg}
\log Z 
= \log {\rm Tr} \,e^{-\beta H_{AB}} 
= \log g +  \frac{c \pi}{6\beta}L - \beta E_0,
\end{equation}
so long as the size of the system $L \gg \beta$ is large\footnote{We
have suppressed the nonuniversal ${\cal O}(T^2)$ and higher-order
finite-$T$ corrections in (\ref{eq:defineg}).}.  Here $H_{AB}$ is the
Hamiltonian for the system with the prescribed boundary conditions,
$c$ is the central charge of the associated CFT, and $E_0$ is the
ground-state energy of $H_{AB}$.

The first two terms in (\ref{eq:defineg}) are universal and determined
by conformal invariance: they depend only on properties of the nearby
critical point.  The third term, $\beta E_0$, is a non-universal
piece: it is sensitive to the details of the theory.  It is clear from
this expression that ${\rm log}(g)$ is the entropy of the system at
zero temperature.  For systems with finite size $L$ the spectrum is
discrete and $g$ is an integer, the ground-state degeneracy ($Z \sim g
e^{-\beta E_0}$).  As $L$ increases the spectrum becomes continuous,
and $g$ is no longer constrained to integer values.

Alternatively, by interchanging the roles of space and time we may
view the boundary conditions $A$ and $B$ as specifying initial and
final boundary states $\ket{A}$ and $\ket{B}$, between which the
system propagates for a time $L$ with periodic boundary conditions on
a spatial circle of circumference $\beta$ \cite{Ishibashi:1988kg,
Cardy:1989ir}.  From this perspective the partition function may be
written
\begin{equation}
Z = \bra{B}e^{-L H_P}\ket{A},
\end{equation}
where $H_P$ is the Hamiltonian of the spatially periodic system.  In
the limit $L/\beta \to \infty$ this becomes
\begin{equation}
\label{gndstate}
Z \sim \braket{A}{0} e^{-L E_0}\braket{0}{B}.
\end{equation}
We thus identify $g \equiv g_A g_{B}= \braket{A}{0}\braket{0}{B}$ and
note that $g$ receives multiplicative contributions from each of the
boundaries, as one would expect (the corresponding contributions to
the entropy are additive).

\section{Perturbative Evaluation of $\mathbf{g}$}\label{sec:calc}

In this section we calculate how $g$ changes under bulk RG flow.
First, however, we discuss some general considerations and summarize
relevant previous results on perturbative flow between critical
points.  Then we explicitly calculate the leading change in $g$.
Finally, we discuss the more general flow.

\subsection{General Considerations and Summary of Previous Results} 
\label{subsec:overview}

Let $S_{\rm{BCFT}}$ be the action for our critical BCFT.  In general,
one may perturb this theory by some combination of relevant bulk
operators $\Phi_i(z,\bar{z})$ and relevant boundary operators
$\Psi_j(x)$.  The resulting theory is described by the action $S$,
\begin{equation}\label{eq:genpert}
S 
= S_{\rm BCFT} + \sum_i \lambda_i \int \Phi_i(z,\bar{z})\, d^2 z 
  + \sum_j \mu_j \int \Psi_j(x)\, dx,
\end{equation}     
and is no longer critical, except perhaps for special values of the
couplings $\lambda_i, \mu_j$.  We can now, at least in principle,
follow the RG flow induced by this perturbation to the nearest RG
fixed point, a theory described by some action $S^\ast$.  If the
theory described by $S^\ast$ is close to the original BCFT then the
resulting flow can be described as a flow in the space of couplings,
and one might hope to understand the system using perturbation theory.
If the perturbing operators $\Phi_i$ and $\Psi_j$ have certain nonzero
operator products with other bulk or boundary operators
$\mathcal{O}_k$ in the theory, then the perturbation
(\ref{eq:genpert}) will induce the flow of those operators
$\mathcal{O}_k$, as well.  (Boundary operators do not induce the flow
of bulk operators, however.)

If the flow from $S_{\rm BCFT}$ to $S^\ast$ involves only boundary
operators, then the endpoint $S^\ast$ describes the same bulk theory
but with different conformally invariant boundary conditions.  If the
flow is induced by a bulk operator then the endpoint will be a new
bulk theory, with conformally invariant boundary conditions consistent
with the new fixed point.  The conformal boundary conditions of
$S^\ast$ may not have any obvious interpretation in terms of the
boundary conditions of $S_{\rm BCFT}$.  This is not an issue in
perturbation theory, however, as one only prescribes boundary
conditions at the original fixed point.

Rather than study the general case, we make two simplifying
assumptions.  First, we assume that the fixed point $S^\ast$ is close
enough to $S_{\rm BCFT}$ that perturbation theory is valid throughout
the flow.  Second, we perturb by only one relevant bulk operator
$\Phi$, and we assume that the operator products of this operator with
the other bulk and boundary operators $\mathcal{O}_k$ are such that
none of their flows are induced by the $\Phi$ perturbation.  With this
second assumption, we can ensure the validity of perturbation theory
by choosing the operator $\Phi$ to be only slightly relevant.

We now recall some important results towards understanding flows
between conformal field theories.  For CFTs without boundary there is
the famous Zamolodchikov $c$-theorem \cite{Zamolodchikov:1986gt}.
This theorem states that the central charges $c$ and $c^*$ of two
fixed points $S_{\rm CFT}$ and $S^*_{\rm CFT}$ must satisfy $c > c^*$
if one can flow from $S_{\rm CFT}$ to $S^*_{\rm CFT}$ along an RG
trajectory.  To prove this theorem Zamolodchikov generalized the
definition of central charge to nonconformal theories by introducing
the so-called $c$-function.  For a conformal field theory, the
$c$-function agrees with the central charge.  Zamolodchikov was able
to show that the $c$-function decreases along RG flow, so the central
charges must decrease.

Cardy and Ludwig \cite{Ludwig:1987gs} addressed this same question in
the regime of perturbation theory.  They perturbed a theory $S_{\rm
CFT}$ by a slightly relevant operator with conformal weights $h =
\bar{h} = 1-y$, where $0 < y \ll 1$, and used perturbative techniques to
investigate the properties of the endpoint $S^*_{\rm CFT}$.  Among
other things, they found the leading-order change $\delta c = c^*-c$
in the central charge to be $\delta c = -y^3/b^2$, where $b$ is the
coefficient in the three-point function of the canonically normalized
perturbing operator.  Because $\delta c$ has a definite sign, this
shows that $c$ decreases in perturbation theory.

For CFTs with boundary the situation is more complicated, because one
must keep track of both the bulk theory and the boundary conditions.
Previous research has focused on purely boundary flows.  Affleck and
Ludwig \cite{Affleck:1991tk} first studied this situation in the
context of the Kondo effect \cite{Affleck:1995ge, PhysRevB.48.7297}.
They originally introduced the quantity $g$ and understood its
physical interpretation.  They studied the change in $g$
perturbatively and found that, under flow generated by a slightly
relevant boundary operator of conformal weight $h = 1-y$, with $0 < y
\ll 1$, $g$ changes by $\delta g = - \pi^2 y^3
g/3b^2$ (again, $b$ is related to the three-point function of the
perturbing boundary operator), and so $g$ decreases in perturbation
theory.  Friedan and Konechny \cite{Friedan:2003yc} have given a
non-perturbative proof of the $g$ theorem.

Assuming $b$ is ${\cal O}(1)$, we can summarize these findings by
saying that, for bulk flow of a boundary-less CFT, $\delta c \sim
-{\cal O}(y^{3})$, and for purely boundary flow of a BCFT, $\delta g
\sim -{\cal O}(y^{3})$.  The quantities $c$ and $g$ decrease under
these flows.  This leaves open the question of how $g$ changes when
the bulk flows between critical points, which is the subject of this
paper.

\subsection{Leading-Order Calculation}

Consider a boundary conformal field theory with action $S_{\rm BCFT}$
that contains a bulk primary operator $\Phi(z,\bar{z})$ with conformal
weights $h = \bar{h} = 1-y$, where $0 < y \ll 1$.  The scaling dimension
$x$ of $\Phi(z,\bar{z})$ is thus $x = h + \bar{h} = 2-2y$.  If we
perturb the action $S_{\rm BCFT}$ by this operator then the resulting
theory is no longer critical, and it has an action $S$ given by
\beq\label{eq:pertaction}
S=S_{\rm BCFT}-a^{-2y} \lambda \int d^{2}z \,\Phi(z,\bar{z}).
\eeq
Here $\lambda$ is the bare coupling and $a$ is a short distance cutoff
required to make $\lambda$ dimensionless.  

Under RG flow the renormalized coupling $\lambda(\ell)$ runs according to
the $\beta$-function equation 
\beq\label{eq:betafunc}
\frac{d\lambda(\ell)}{d\log(\ell)}
=\beta(\lambda) = 2y \lambda - \pi b \lambda^2 + \cdots,
\eeq
and so there are fixed points at $\lambda=0$ and $\lambda =
\lambda^{*}\equiv 2y/\pi b$.  We take $\ell$ to be the length scale
at which the theory is defined.  The constant $b$, which we take to be
${\cal O}(1)$, is related to the coefficient of the bulk three-point
function of $\Phi$:\footnote{$\beta(\lambda)$ is determined by the
bulk theory alone, so this three-point correlation function is
computed in the theory without boundary.  We normalize $\Phi$ to have
the standard two-point function
$\langle\Phi(z_1,\bar{z}_1)\Phi(z_2,\bar{z}_2)\rangle = |z_{1} -
z_{2}|^{-x}$.}
\begin{equation}
\label{3}
\langle\Phi(z_1,\bar{z}_1)\Phi(z_2,\bar{z}_2)\Phi(z_3,\bar{z}_3)\rangle 
=
\frac{-b}{|z_{12}|^x|z_{23}|^x|z_{31}|^x}.
\end{equation}
Note that in writing (\ref{eq:betafunc}) we have assumed that the
$\Phi(z,\bar{z})$ perturbation does not induce the flow of any other
relevant bulk operators.  Furthermore, we will assume that no boundary
flows are induced.  We relax these assumptions and discuss the
more general flow in the next section.

We wish to calculate the change in $g$ between the two fixed points,
when $\lambda^{*} \sim y \ll 1$ is sufficiently small that
perturbation theory is valid throughout the flow.  In general one must
regulate a perturbative calculation by considering a finite-size
system at non-zero temperature.  We encounter no divergences with
length, so we are free to put our system on an infinite half-cylinder.
The inverse temperature is then encoded in the circumference $\beta$,
and the single conformal boundary condition we label by $k$.  (We
shall therefore be studying the boundary entropy $g_k$ associated to
$k$.)

In order to compute the change $\delta g/g$, consider 
\beq\label{eq:partrat}
\log Z - \log Z_0
= \frac{Z}{Z_{0}} -1 + \mathcal{O}(\lambda^2) 
= a^{-2y}\lambda
   \int d^{2}w \,\langle \Phi(w,\bar{w})\rangle_{\rm CYL} + \cdots,
\eeq
where $Z$ ($Z_{0}$) is the partition function of the (un)perturbed
system.  Note that the leading-order correction in (\ref{eq:partrat}),
which we shall hereafter denote by $Z_1$, involves the one-point
function of the bulk operator $\Phi$.  If our theory were on a
manifold without boundary then the operator $\Phi$ could be chosen to
have vanishing one-point function, and then the leading correction
would be given by a two-point function.  Similarly, if we were to
perturb our BCFT by a boundary operator, then the one-point function
of that operator along the boundary could be chosen to vanish.  But
the one-point function of a bulk operator on a manifold with boundary
does not vanish in general, and, in fact, it depends on the boundary
conditions.

From (\ref{eq:defineg}) we see that $Z_1$ could contain three
contributions: the term $\log(1+\delta g/g) \approx \delta g/g$ (to
lowest order) that interests us, a piece that depends on the central
charge, and a ground-state energy correction.  To the order that we
are working the central charges of the two fixed points $S_{\rm BCFT}$
and $S^*$ agree, and so the corresponding corrections will not appear
in our computation.  Thus to isolate the term $\delta g/g$ we need
only identify and discard the ground-state energy renormalizations.
This is easy because such contributions scale linearly with the
inverse temperature $\beta$.

The one-point function on the half cylinder can be computed by
conformal transformation from the familiar result for the upper half plane.
The one-point function on the upper half plane with standard
coordinate $z$ is constrained by conformal invariance to be of the
form
\begin{equation}\label{eq:UHPcorrelator}
\langle\Phi(z,\bar{z})\rangle_{\rm UHP}
= \frac{A_\Phi^k}{|2\, {\rm Im}\, z|^x},
\end{equation}
where $A_\Phi^k$ depends on both the particular field $\Phi$ and the
choice of conformal boundary condition $k$.  Because the coefficient
$A_\Phi^k$ has no definite sign, we will see that $g$ can increase
during bulk RG flow.

Let us
coordinatize the cylinder by $w = \tau + i \sigma$; the imaginary
direction $\sigma$ winds around the compact circle, and the real
direction $\tau$ measures distance from the boundary.  Then a suitable
conformal mapping to the half plane is $z = i \tanh (\pi w/\beta)$.
We can now compute
\begin{equation}
\langle\Phi(w,\bar{w})\rangle_{\rm CYL}
= \left|\frac{dz}{dw}\right|^x
  \langle\Phi(z,\bar{z})\rangle_{\rm UHP}
= \frac{A_\Phi^k}
       {\left(\frac{\beta}{\pi}\sinh(\frac{2\pi}{\beta}\tau)\right)^x}.
\end{equation}

The leading contribution $Z_1$ is thus
\beq
\label{Z1}
Z_1 =
a^{-2y} \lambda \int_0^{\beta}d\sigma 
\int_{a}^{\infty} d\tau 
\frac{A_{\Phi}^{k}}
     {(\frac{\beta}{\pi} \sinh( \frac{2 \pi}{\beta} \tau) )^{2-2y}}.
\eeq
After performing the integral over the compact $\sigma$ coordinate and
changing variables to $r = \exp(4\pi\tau/\beta)$ we are left with
\beq
Z_1  = \pi \lambda A_\Phi^k \bigg(\frac{\beta}{2\pi a}\bigg)^{2y} 
              \int_{a'}^{\infty} \frac{dr}{r^y (r-1)^{2-2y}},
\eeq
where $a' =e^{\frac{4 \pi a}{\beta}}$.  The integral diverges in the
limit $a' \to 1$ of small cutoff ($a \to 0$).  In order to understand
the nature of this divergence we integrate by parts to extract it from
the integral. 

Integrating by parts once gives
\beq\label{eq:firstintpart}
Z_1
= -\pi\lambda A_{\Phi}^{k}\frac{1}{1-2y}\left(\frac{\beta}{2\pi a}\right)^{2y}
  \left[\left.\frac{1}{r^y(r-1)^{1-2y}} \right|_{a'}^{\infty}
       +y\int_{a'}^{\infty} \frac{dr}{r^{1+y} (r-1)^{1-2y}} \right].
\eeq
The integral in (\ref{eq:firstintpart}) is now convergent as $a^\prime
\to 1$.  However, it diverges at $y = 0$, and our ultimate goal is a
perturbative calculation in $y$.  To remedy this we integrate by parts
a second time:
\beq
\label{bp2}
Z_1
= -\pi\lambda A_{\Phi}^{k}\frac{1}{1-2y} \left(\frac{\beta}{2\pi a}\right)^{2y}
    \left[\left.\frac{(r-1)^{2y}}{r^y}
       \frac{3r - 1}{2r(r-1)} \right|_{a'}^{\infty}
       +\frac{1+y}{2}\int_{a'}^{\infty}\frac{(r-1)^{2y}}{r^{2+y}}dr \right].
\eeq

The prefactor $\lambda (\beta/a)^{2y}$ will eventually be expressed in
terms of the renormalized coupling, so for now we concentrate on the
$y$ dependence inside the brackets.  The integral in (\ref{bp2}) is
convergent as $a'\to 1$ and $y \to 0$, so we may expand it in
powers of $y$. To lowest order, we find
\beq
\label{bp3}
Z_1
= \pi \lambda A_{\Phi}^{k} \left(\frac{\beta}{2 \pi a}\right)^{2y}
   \left[\left({\beta \over 4 \pi a}\right)^{1-2y}-\frac{1}{2}
    + \mathcal{O}\bigg(y, {a \over \beta}\bigg) \right].
\eeq
The first term is simply a ground-state energy correction; this we
discard in our calculation of $\delta g/g$.  The second term is the
leading contribution to the change in $g$.  

In order to determine how $g$ changes between the two fixed points we
must rewrite $\delta g/g$ in terms of the renormalized coupling
$\lambda(\beta)$ (we take $\beta$ to be the length scale $\ell$ at which 
the coupling is defined).  The solution to the RG equation
(\ref{eq:betafunc}) is
\beq
\label{renorm}
\lambda\left(\frac{\beta}{a}\right)^{2y} 
= \frac{\lambda(\beta)}{1-\frac{\lambda(\beta)}
        {\lambda^{*}}(1-(\frac{a}{\beta})^{2y})},
\eeq
where $\lambda = \lambda(a)$ is the coupling evaluated at the UV
cutoff.  In the regime $\lambda (\beta)\ll \lambda^{*}$ we can
approximate $\lambda(\beta) \approx \lambda(\beta/a)^{2y}$ and write
$\delta g/g$ in terms of the renormalized coupling $\lambda(\beta)$:
\begin{equation}
\frac{\delta g}{g}
 = -\frac{\pi}{2} \lambda(\beta) A_\Phi^k + \mathcal{O}(y^2).
\end{equation}
So long as the new fixed point $\lambda^{\ast} \ll 1$ is close to zero
and perturbation theory is valid, this last expression should remain
true throughout the entire flow.  In particular, it will remain true
in the low-temperature limit $\beta \to \infty$, as the coupling
approaches the new critical point $\lambda^{\ast} = 2y/\pi b$.

We find
\beq
\label{change}
\frac{\delta g}{g} = -\frac{y}{b}A_{\Phi}^{k}+\mathcal{O}(y^2).
\eeq
Thus $g$ can either increase or decrease, depending upon the relative
signs of $A_{\Phi}^{k}$ and $b$.  Note that the sign of
$A_\Phi^k$ depends on both the choice of bulk operator $\Phi$ and the
choice of boundary condition $k$; it is not determined by the bulk
theory.

\subsection{Induced Flows and Higher-Order Terms}

In our leading-order calculation we assumed that the $\Phi$
perturbation did not induce the flow of any other operators.  However,
bulk-induced boundary operators appear frequently
\cite{Fredenhagen:2006dn,Baumgartl:2007an,Fredenhagen:2007rx,
Green:2006ku}, and in many cases they are the dominant effect.  These
flows generally reduce $g$ to the lowest available value.  The bulk
perturbation, meanwhile, may increase $g$ for some boundary conditions
and decrease it for others.  

Suppose the bulk perturbation (\ref{eq:pertaction}) induces the flow
of a boundary operator $\Psi$ of scaling dimension $1-x$, with $x \ll
1$.  The $\beta$-function for the coupling $\mu$ of the operator
$\Psi$ is \cite{Fredenhagen:2006dn}
\beq
\beta(\mu) = x \mu + B_{\Phi \Psi} \lambda - b_\Psi \mu^2 + \ldots,
\eeq
where $B_{\phi \psi}$ is the coefficient of the bulk-boundary OPE of
$\Phi$ and $\Psi$, and $b_\Psi$ is the coefficient of the boundary
three-point function.  The behavior of $g$ now depends on the relative
magnitudes of $x$ and $y$.  For example, if $y^{\frac{1}{3}} \ll x \ll
1$ then the boundary RG flow will dominate, and $g$ will decrease.  In
regimes where the purely boundary and purely bulk effects are
competitive, one would also need to include the contribution from the
bulk-boundary correlator.

Above we found that bulk perturbations with $A_{\Phi}^{k}=0$ leave $g$
unchanged to leading order in $y$.  To better understand the effects
of such perturbations we would need to include higher-order
corrections.  In cases where there is no induced flow the next
correction arises from the bulk two-point function.  This is more
difficult to address, because the two-point function of a primary
operator on a manifold with boundary is only fixed by conformal
invariance up to a function (rather than just a constant), which
depends upon the specifics of the theory.  We did not compute this
contribution.

\subsection{Marginal Operators}\label{sec:marginal}

Exactly marginal operators describe a continuous family of conformal
field theories, labeled by the coupling of the operator.
Understanding the space of such theories is different from
understanding the behavior of a single theory under RG flow.
Nevertheless, the exactly marginal case can be viewed as a limit of RG
flow between nearby fixed points.  In this spirit we may adapt the
above analysis to the exactly marginal case.

We shall again assume that the deformation induces no boundary flows.
This requires the existence of a continuous family of conformal
boundary conditions, also labeled by the coupling.  If no such family
of boundary conditions exists then the deformation will necessarily
induce boundary flow.  For a more detailed study of exactly marginal
operators, see
\cite{Fredenhagen:2006dn,Baumgartl:2007an,Fredenhagen:2007rx}.

To understand the behavior of $g$ in  the marginal case we simply evaluate 
(\ref{Z1}) at $y=0$.  The calculation is easier than before, as the integral 
can be worked out explicitly.  The result is
\beq
Z_1 
= \frac{\pi}{2} \lambda A_\Phi^k \left(\frac{\beta}{4\pi a}-1\right)
  + \mathcal{O}\left(\frac{a}{\beta}\right).
\eeq
As before, the first term is a correction to the ground state energy, and the
second term is the contribution to $\delta g/g$.  So, for an exactly marginal
deformation,
\beq
\frac{\delta g}{g} = -\frac{\pi}{2} \lambda A_\Phi^k + \mathcal{O}(\lambda^2).
\eeq
Evidently, marginal deformations can also serve to either
increase or decrease $g$.

A peculiar application of this was already known from
\cite{Recknagel:1998ih,Sen:1999mh}, where it was found that one can
reverse the effects of a relevant boundary flow by a series of
marginal deformations.  The physics involved is simply that of a free
boson on a circle.  The relevant boundary flow in question changes the
boundary conditions from Neumann to Dirichlet.  However, if the size
of the circle is at the self-dual point then there exists a marginal
boundary deformation that changes the boundary conditions back to
Neumann from Dirichlet.  Furthermore, the bulk theory contains an
exactly marginal operator that changes the size of the circle.  It is
now clear how to reverse the effects of the initial relevant boundary
flow.  Simply deform the size of the circle to the self-dual point,
return the boundary conditions to Neumann, and then restore the circle
to its original size.  For Neumann boundary conditions $g$ is
proportional to the radius of the circle, so during this procedure the
marginal bulk deformation increases the value of $g$.

\section{Minimal Models}\label{sec:mm}

Now we provide a class of examples in which $g$ increases during bulk
RG flow.  The BCFTs of interest are the minimal models with boundary.

The minimal models are the unitary rational conformal field theories
with central charge $c = 1 - \frac{6}{m(m+1)} < 1$, where $m \geq 3$.
Their (primary) operator content is parameterized by two integers
$(r,s)$, with $1 \le r \le m-1$ and $1 \le s \le m$, modulo the
conformal grid symmetry $(r,s) \sim (m-r, m+1-s)$.  The conformal
weights of the operator $\Phi_{(r,s)}$ are
\begin{eqnarray}
\label{h}
h_{m}(r,s) = \overline{h}_{m}(r,s) 
= \frac{((m+1)r-ms)^{2}-1}{4m(m+1)}.
\end{eqnarray}  
For a BCFT there is a one-to-one correspondence between primary bulk
operators and conformally invariant boundary conditions.

We are interested in a specific class of flows, first understood
perturbatively by Zamolodchikov \cite{Zamolodchikov:1987ti}.  When a
minimal model at level $m$ is deformed by its least relevant operator,
$\Phi_{(1,3)}$, the theory flows to the level $m-1$ minimal model.  In
the limit of large $m$ these theories cluster near $c = 1$, and these
RG flows can be understood using a perturbative expansion in $m^{-1}$.  
Note that $\delta c \sim
1/m^{3}$ between nearby minimal models, consistent with the
assumptions of our general calculation in section~\ref{sec:calc}.

We will apply our perturbative result (\ref{change}) to estimate the
resulting change in $g$.  However, our previous result was derived
under the assumption that the bulk perturbation does not induce any
boundary flows.  We were unable to calculate the general bulk-boundary
correlator, and so this is an assumption that we could not explicitly
verify.  Nevertheless, perturbative boundary flows would contribute only
subleading corrections, so we need only worry about large boundary 
flows that make $\mathcal{O}(1)$ contributions. 
The minimal models are so simple that one can explicitly
calculate $g$ for a given choice of boundary conditions.  We find
agreement between our perturbative calculation of $\delta g$ and the
exact result (to leading order) when the change in $g$ is small enough
that it can be understood in perturbation theory.  We shall return to
this point below.

First we review some useful facts about boundary conditions in minimal
models.  In section~\ref{sec:g} we saw that a boundary condition can
be thought of as specifying a boundary state $\ket{B}$.  A
particularly simple basis of such states was discovered by Cardy and
Lewellen \cite{Cardy:1991tv}, and the basis states $\ket{B_{(r,s)}}$
are known as Cardy states.  They carry a label $(r,s)$ just like the
primary operators.  These boundary states $\ket{B_{(r,s)}}$ have the
property that their overlap $\braket{\Phi_{(a,b)}}{B_{(r,s)}}$ with
the state $\ket{\Phi_{(a,b)}}$ (that corresponds to the primary
operator $\Phi_{(a,b)}$ in the standard state-operator correspondence)
is given by
\begin{equation}\label{eq:cardyoverlap}
\braket{\Phi_{(a,b)}}{B_{(r,s)}}
= \frac{S_{(r,s)}^{(a,b)}}{\sqrt{S_{(1,1)}^{(a,b)}}},
\end{equation}
where $S$ is the modular S-matrix
\beq\label{S}
S_{(r,s)(r',s')} 
= (-1)^{1+rs'+sr'}\sqrt{\frac{8}{m(m+1)}} 
   \sin\left[\pi \left(\frac{m+1}{m}\right)r r'\right]
    \sin\left[\pi \left(\frac{m}{m+1}\right) s s'\right].
\eeq

From the expression (\ref{eq:cardyoverlap}) we may compute all of the
quantities that are of interest to us.  For example, we saw in
section~\ref{sec:g} that the boundary entropy $g_{(r,s)}$ of the
boundary condition $(r,s)$ is given by $\braket{0}{B_{(r,s)}}$.  For
unitary CFTs the vacuum state $\ket{0}$ corresponds to the identity
operator $1 = \Phi_{(1,1)}$, so we find
\begin{equation}\label{eq:gbraket}
g_{(r,s)}
= \braket{\Phi_{(1,1)}}{B_{(r,s)}}.
\end{equation}
Similarly, the properly normalized one-point function of the operator
$\Phi_{(a,b)}$ on the upper half plane with boundary condition $(r,s)$
can be computed in terms of boundary states.  In the notation of
(\ref{eq:UHPcorrelator}) it is
\begin{equation}\label{eq:minimalA}
A_{(a,b)}^{(r,s)} 
= \frac{\braket{\Phi_{(a,b)}}{B_{(r,s)}}}{\braket{0}{B_{(r,s)}}}.
\end{equation}

We can now begin to understand the flows induced by the $\Phi_{(1,3)}$
operator.  The conformal weights of $\Phi_{(1,3)}$ are given by
(\ref{h}).  To leading order in $m^{-1}$ this is $h = \bar{h} =
1-2/m$, so we have $y = 2/m$.  The coefficient $A_{(1,3)}^{(r,s)}$ of
the one-point function is given in (\ref{eq:minimalA}).  To leading
order in $m^{-1}$, and assuming $r, s \ll m$, $A^{(r,s)}_{(1,3)} =
\sqrt{3}$.  The three-point correlation function was calculated in
\cite{Dotsenko:1984nm}, and to leading order the coefficient is
$b=-4/\sqrt{3}$.  Putting all this together in (\ref{change}) yields
\beq\label{eq:ourmmgchange}
\frac{\delta g}{g} = -\frac{y}{b}A_{(1,3)}^{(r,s)} = \frac{3}{2m}.
\eeq
This is positive, and so we conclude that $g$ always increases in the regime
$r, s \ll m$.  Although (\ref{eq:ourmmgchange}) was derived under the
assumption of no induced boundary flow, any perturbative boundary flow
would contribute a subleading correction at order
$\mathcal{O}(m^{-3})$.

To understand the matter more fully, we use our knowledge of minimal
models to calculate the change in $g$ when the initial theory at level
$m$ has Cardy boundary condition $(r,s)$ and the final theory at level
$m-1$ has $(r^*,s^*)$.  Using (\ref{eq:gbraket}) we find
\beq
g_{(r,s)}(m) 
= \left[\frac{8}{m(m+1)}\right]^{\frac{1}{4}}
   \frac{\sin(\frac{\pi r}{m}) \sin(\frac{\pi s}{m+1})}
        {\sqrt{\sin( \frac{\pi}{m})\sin(\frac{\pi}{m+1})}}.
\eeq
Expanding to $\mathcal{O}(m^{-1})$, and assuming $r,s \ll m$, gives
\beq
\label{mmgchange}
\frac{\delta g}{g} 
= \log\left[\frac{g_{(r,s)}(m-1)}{g_{(r^*,s^*)}(m)}\right] 
= \log\left(\frac{rs}{r^*s^*}\right)+\frac{3}{2m}.
\eeq
We see that the change $\delta g$ contains two pieces: an order
$\mathcal{O}(1)$ logarithm and an $\mathcal{O}(m^{-1})$
correction.  The $\mathcal{O}(m^{-1})$ perturbation agrees with our
calculation (\ref{eq:ourmmgchange}).  

Equations (\ref{eq:ourmmgchange}) and (\ref{mmgchange}) provide a
constraint on the allowed perturbative flows between nearby boundary
minimal models.  Specifically, given an initial boundary Cardy state
$(r,s)$, an endpoint Cardy state $(r^*,s^*)$ must satisfy $rs =
r^*s^*$.  If the endpoint boundary condition is instead a linear
combination of Cardy states, then cancellations must occur to agree
with the perturbative result (\ref{eq:ourmmgchange}).  

The agreement betweeen (\ref{eq:ourmmgchange}) and (\ref{mmgchange})
is somewhat remarkable.  A similar analysis, for purely boundary flows
in minimal models, was performed in \cite{Recknagel:2000ri}.  There it
was shown that this simple type of analysis fails for most boundary
conditions.  In particular, matching the perturbative and exact
results could not be accomplished with a simple flow between Cardy
states, but rather required the end point to be some linear
combination.  The agreement of our results for the Cardy states
satisfying $r^*s^* = rs$ is consistent with the existence of such
flows, which would increase $g$.

One difficulty in understanding boundary flows lies in the fact that
the initial and final bulk theories are different, so boundary
conditions labeled by the same integer pairs will have different
meaning before and after the bulk RG flow.  Thus it is not clear what
endpoint boundary condition results from a given initial boundary
condition, even when there is no induced boundary flow.  More
prosaically, the endpoint $m-1$ minimal model contains fewer primary
operators than the initial $m$ theory, and so it admits
correspondingly fewer conformally invariant boundary conditions.

We suspect that the flow from $(1,1)$ to $(1^*,1^*)$ exists.  As
mentioned above, Zamolodchikov showed \cite{Zamolodchikov:1987ti} that
the $\Phi_{(1,3)}$ operator induces flow between nearby minimal
models.  This should remain true in the presence of a boundary so long
as the initial boundary conditions can smoothly flow into some
boundary condition of the resulting theory.  The
$(1,1)$ boundary state has the lowest $g$ among Cardy states, so it is
unlikely that it induces boundary flow.  Furthermore, it is known that, in the bulk 
flow induced by $\Phi_{(1,3)}$, the primary operators $\Phi_{nn}$ of the $m$ theory
flow to the $\Phi_{nn}$ operators of the $m-1$ theory
\cite{Zamolodchikov:1987ti}.  Thus, noting the
one-to-one correspondence between boundary states and bulk primary
operators, it seems reasonable that the Cardy
state $(1,1)$ should flow to the corresponding boundary condition in
the $m-1$ theory.

\section{Interpretations and Applications}\label{sec:interp}

\subsection{String Theory}
Renormalization group flow in the bulk of a two-dimensional field
theory is often used to understand closed string tachyon condensation.
In this context the relevant operator is the vertex operator for a
negative-mass state in the spacetime description.  

RG flow of the worldsheet field theory can approximate time evolution
of the target space theory
\cite{Polchinski:1989fn,Cooper:1991vg,Schmidhuber:1994bv,Freedman:2005wx}, 
but in general the two are different.  In particular, the RG flow
yields first-order equations of motion, while the time evolution is
governed by second-order equations.  RG flow becomes a good
approximation in supercritical theories with large, negative dilaton
time derivations
\cite{Polchinski:1989fn,Cooper:1991vg,Schmidhuber:1994bv,Freedman:2005wx}.
In such cases the friction term dominates the time evolution, yielding
approximately first-order behavior (much like slow-roll inflation).

In other cases RG flow yields the same endpoint as time evolution.
This happens for tachyon condensation processes that are localized in
space, where energy density can escape to infinity.  With decays to
``nothing," where entire regions of spacetime disappear, RG and time
evolution seem to give qualitatively similar results.  In these
scenarios all of the closed string states become exponentially massive
(including the graviton), and it is in this way that the spacetime
description is lost.  Here the agreement between the two
approaches might be due to the graviton mass, which impedes
backreaction.

The inclusion of boundaries on the worldsheet now describes a
spacetime containing D-branes.  The tension of these branes is given
by $g$ \cite{Harvey:1999gq}.  The behavior of D-branes under closed
string tachyon condensation is potentially very interesting, as they
may react to the condensate in a qualitatively differently way from
the usual string states.  If they survive into regions where string
states become lifted, then this might signal a breakdown in the
unitarity of the spacetime theory.  Just this phenomenon has been
argued to occur in the $c=1$ matrix model \cite{Karczmarek:2004ph}.
While the underlying theory remains unitary, it is still disturbing to
lose unitarity in spacetime.

In other contexts \cite{Green:2006ku,Gaberdiel:2007us} it has been
shown that this does not occur: the branes gain exponentially large
masses ($g \to \infty$) \cite{Green:2006ku} or completely decouple ($g
\rightarrow 0$) \cite{Gaberdiel:2007us}.  Both would be consistent
with our expectations from the closed string sector.  The decoupling
case is similar to open string tachyon condensation processes, where
the brane disappears from the spacetime description.  This is
distinguished from the D-brane simply becoming light by the absence of
couplings to any closed string modes (term-wise vanishing of the
open-string partition function, as opposed to cancellations between
separate contributions).  To preserve spacetime unitarity one would
prefer the branes to become massive or decouple, rather than persist
unaffected.

Our results suggest that the brane tension is sensitive to the tachyon
condensation even if the tachyon does not modify the boundary
conditions.  Closed string tachyon condensation alone can cause the
brane to become extremely massive.  We have not shown that this is the
generic outcome (although it is suggested in \cite{Green:2006ku}).
However, the sensitivity of the tension to the bulk RG makes it
unlikely for the brane to be unaffected by a closed string tachyon
condensate.

One further motivation for this work was to resolve some confusions
associated with \cite{Green:2006ku}.  The definition of ``mass" used
there is not equivalent to $g$, so it was unclear whether $g$ should
increase along these flows.  This work is consistent with the
interpretation that the branes do have growing masses, as the results
in both cases depend on the sign of $A_{\Phi}^{k}$ in the same way.

Of course, string theory is more than just CFT on a disk; it includes
a sum over all metrics and topologies of the worldsheet, as well.
Taking this into account, \cite{Keller:2007nd} determined the
backreaction of a brane on the \emph{spacetime} geometry.  The leading
effect, due to the Fischler-Susskind mechanism, is a modification of
the beta function for the bulk coupling:
\beq
\frac{d \lambda}{d \log(\ell)}
= \beta(\lambda)
= 2y\lambda + g_s\frac{g}{\pi} A_{\Phi}^{k}- \pi b \lambda^2,
\eeq
where $g_{s}$ is the string coupling.  If one can stabilize the
dilaton such that $g_{s} \ll y^2/bgA_\Phi^k$ then our perturbative
analysis holds, though with a slighly modified fixed point, yielding
\beq
\frac{\delta g}{g} 
= -\frac{y}{b}A_{\Phi}^{k}
  -g_s\frac{g}{4y} (A_{\Phi}^{k})^2  
  +\mathcal{O}(y^2).
\eeq
We see that the backreaction always decreases $g$, in this limit.
This is consistent with the observations \cite{Green:2006ku,
Keller:2007nd, Gaberdiel:2007us} that induced boundary flows and
backreactions tend to minimize the mass of branes.

\subsection{Condensed Matter}

Two-dimensional boundary conformal field theories have been used to
understand the multi-channel Kondo effect
\cite{Affleck:1990iv,PhysRevB.48.7297,Ludwig:1994nf} (see
\cite{Affleck:1995ge} for a review).  In fact, it is in this context
that Affleck and Ludwig originally defined $g$ \cite{Affleck:1991tk}.
Here one is trying to understand the role of impurities that are
coupled via a spin interaction to a system of free fermions.  The
problem is reduced to 1+1 dimensions by concentrating on $s$-wave or
radial scattering of the electrons from the impurity.  In the UV, the
coupling of the impurity to the electrons acts as a relevant boundary
operator perturbing the free fermion BCFT.  In the IR, at the 
strongly-coupled fixed point, the theory is described by free fermions with a
new, nontrivial boundary condition.  This approach is particularly
powerful in the over-screened case, where the appropriate boundary
condition has no simple description in terms of free fermions.
Nevertheless, one can deduce the correct boundary theory from the
fusion rules.

In this context $\log(g)$ is the impurity entropy.  When there is a
suitable free fermion description (the critical and under-screened
cases), $g$ agrees with the ground-state degeneracy, which can be
determined independently.  In all cases $g$ is lower at the 
strong-coupling fixed point (IR) than at the free fixed point (UV).  This
follows directly from the $g$ theorem.

In this language our result is simply a statement about the effects
that bulk interactions can have on the impurity entropy.  While the free
fermion description might be a good approximation for some systems, this
is not the case in general.  Our results show that the impurity entropy
can increase or decrease as we lower the temperature of the system,
due only to interactions between the fermions.

Further applications would be interesting.  Recently
\cite{Fendley:2006gr}, a connection has been made between the boundary
entropy and the universal part of the entanglement entropy
\cite{Kitaev:2005dm, LevinWen} of a system exhibiting a topological
phase .  Both quantities are related to a particular entry of the
modular S-matrix of an associated CFT (for further elaboration, please see 
\cite{topphase}).  We would like to make more direct contact between $g$ and
the physics of topological phases.

\section{Conclusion}

We have studied the effect of bulk RG flow on the boundary entropy of
a two-dimensional BCFT when the flow is due to perturbation by a
single (slightly) relevant primary operator of scaling dimension
$2-2y$.  To leading order the change in $g$ was found to be $\delta
g/g = -A y/b$, where $A$ is the coefficient of the one-point function
and $b$ the coefficient of the three-point function.  This expression
has no preferred sign, and so $g$ can either increase or decrease,
depending upon the choice of perturbing operator and boundary
conditions.  We saw an explicit realization of this in the minimal
models, where the flows induced by the $\Phi_{(1,3)}$ operator can
increase $g$ for certain boundary conditions.  All of this is in stark
contrast to the behavior of $g$ under purely boundary perturbations,
for which the $g$ theorem guarantees that $g$ will decrease.  Our
result has applications to closed string tachyon condensation and the
multi-channel Kondo effect.

\acknowledgments
We would like to thank Eduardo Fradkin, Matt Headrick, Shamit Kachru,
Eun-Ah Kim, Albion Lawrence, and Eva Silverstein for helpful
discussions.  We are supported in part by the DOE under contract
DE-AC03-76SF00515 and by the NSF under contract 9870115.  DG is also
supported by a NSERC Postgraduate fellowship and a Mellam Family
Foundation Fellowship.

\bibliography{gbib22}

\end{document}